# Neural Substitute Solver for Efficient Edge Inference of Power Electronic Hybrid Dynamics

Jialin Zheng, *Member, IEEE*, Haoyu Wang, *Student Member, IEEE*, Yangbin Zeng, *Member, IEEE*, Han Xu, *Student Member, IEEE*, Di Mou, *Member, IEEE*, Hong Li, *Senior Member, IEEE*, Sergio Vazquez, *Fellow, IEEE*, Leopoldo G. Franquelo, *Life Fellow, IEEE*

*Abstract*—Advancing the dynamics inference of power electronic systems (PES) to the real-time edge-side holds transformative potential for testing, control, and monitoring. However, efficiently inferring the inherent hybrid continuous-discrete dynamics on resource-constrained edge hardware remains a significant challenge. This letter proposes a neural substitute solver (NSS) approach, which is a neural-network-based framework aimed at rapid accurate inference with significantly reduced computational costs. Specifically, NSS leverages lightweight neural networks to substitute time-consuming matrix operation and high-order numerical integration steps in traditional solvers, which transforms sequential bottlenecks into highly parallel operation suitable for edge hardware. Experimental validation on a multi-stage DC-DC converter demonstrates that NSS achieves 23x speedup and 60% hardware resource reduction compared to traditional solvers, paving the way for deploying edge inference of high-fidelity PES dynamics.

## I. Introduction

Real-time dynamics inference of power electronic systems (PES) is the basis for a wide range of model-based design, test and control developments, and plays a central role in a wide range of advanced applications such as hardware-in-the-loop (HIL) testing, digital twins (DT) and model predictive control (MPC) [1], [2]. Driven by the need for improved system responsiveness, data privacy, and local autonomy, advancing high-fidelity dynamics inference from centralized platforms to edge devices near physical systems has become a clear trend [3].

However, achieving efficient inference on resource-constrained edge devices faces significant challenges from the inherent hybrid dynamics of PES [4]. Discrete switching events caused by digital control frequently alter circuit topology, and are coupled with continuous energy-storage elements to regulate energy flows [5]. Therefore, the inference needs a numerical solver that not only accurately detects measurement events and updates the system model, but also precisely integrates the continuous dynamics within the event interval.

Conventional numerical solvers exhibit limitations in addressing above challenges. Fixed-step methods can ensure event detection accuracy by extremely small step sizes and facilitate parallelization on hardware like FPGAs in low-order algorithms [6]. However, it incurs massive computational costs and stringent real-time constraints, demanding substantial computing resources. Furthermore, it may necessitate system model updates, including expensive matrix operation like inversion at each step. Conversely, variable-step methods employ higher-order integration and event-driven schemes to improve per-step efficiency and accuracy while reducing the model update frequency [4], [5]. However, in real-time edge inference, highly variable step sizes and fluctuating computation time make it difficult to meet strict timing constraints. Additionally, the sequential nature of high-order algorithms hinders effective parallelization on FPGAs and online matrix operation [4]. Therefore, existing methods present a trade-off among real-time capability, computing accuracy, and edge resource consumption.

While model reduction can accelerate inference speed for specific converters [7], an effective universal methodology is still lacking for efficient edge dynamics inference of PES. To bridge this gap, this letter proposes a Neural Substitute Solver (NSS), aiming to enhance efficiency by simultaneously improving inference speed and reducing computational costs. By leveraging the universal approximation capabilities of neural networks (NN), NSS employs a dual-network architecture: 1) one substitutes the online system model update process due to discrete events, and 2) the other enhances a base numerical integrator by approximating higher-order error terms of continuous state. The main contributions of this paper are: 1) a generic NN substitute solver framework for hybrid PES dynamics inference; 2) a lightweight dual-NN solver design and its sequential training scheme; 3) A highly parallelized neural processing units for accelerating the NSS on edge devices.

The rest of the letter is organized as follows. Section II describes the problem setup, Section III introduces the NSS and its implementation, and Section IV discusses the experimental results. Finally, Section V concludes the letter.

## II. Background: Hybrid Dynamics Inference of PES

### A. Hybrid Dynamics of PES

The dynamics of PES are typically described by general ordinary differential equations (ODE):

$$\dot{x}(t) = f(t, x(t), u(t)), \quad t \in \mathcal{T} \tag{1}$$

where $x \in \mathbb{R}^n$ is the state vector (e.g., capacitor voltages and inductor currents), $u \in \mathbb{R}^m$ is the input vector, and $t$ denotes time. The hybrid nature of PES arises from discrete switching events, whose precise time is critical for simulation accuracy, as illustrated in Fig. 1(a). In the $k$-th interval between successive events, the system dynamics are determined by a set of switching events, behaving as a linear time-invariant system:

$$\dot{x}(t) = A_k x(t) + B_k u(t) \tag{2}$$

where system matrices $A_k$ and $B_k$ can be modeled as,

Jialin Zheng and Haoyu Wang contributed equally to this work.
This paper was supported by the Fundamental Research Funds for the Central Universities under Grant 2024ZYGXZR073 and supported by State Key Laboratory of Power System Operation and Control under Grant SKLD24KM27. *(Corresponding author: Yangbin Zeng.)*

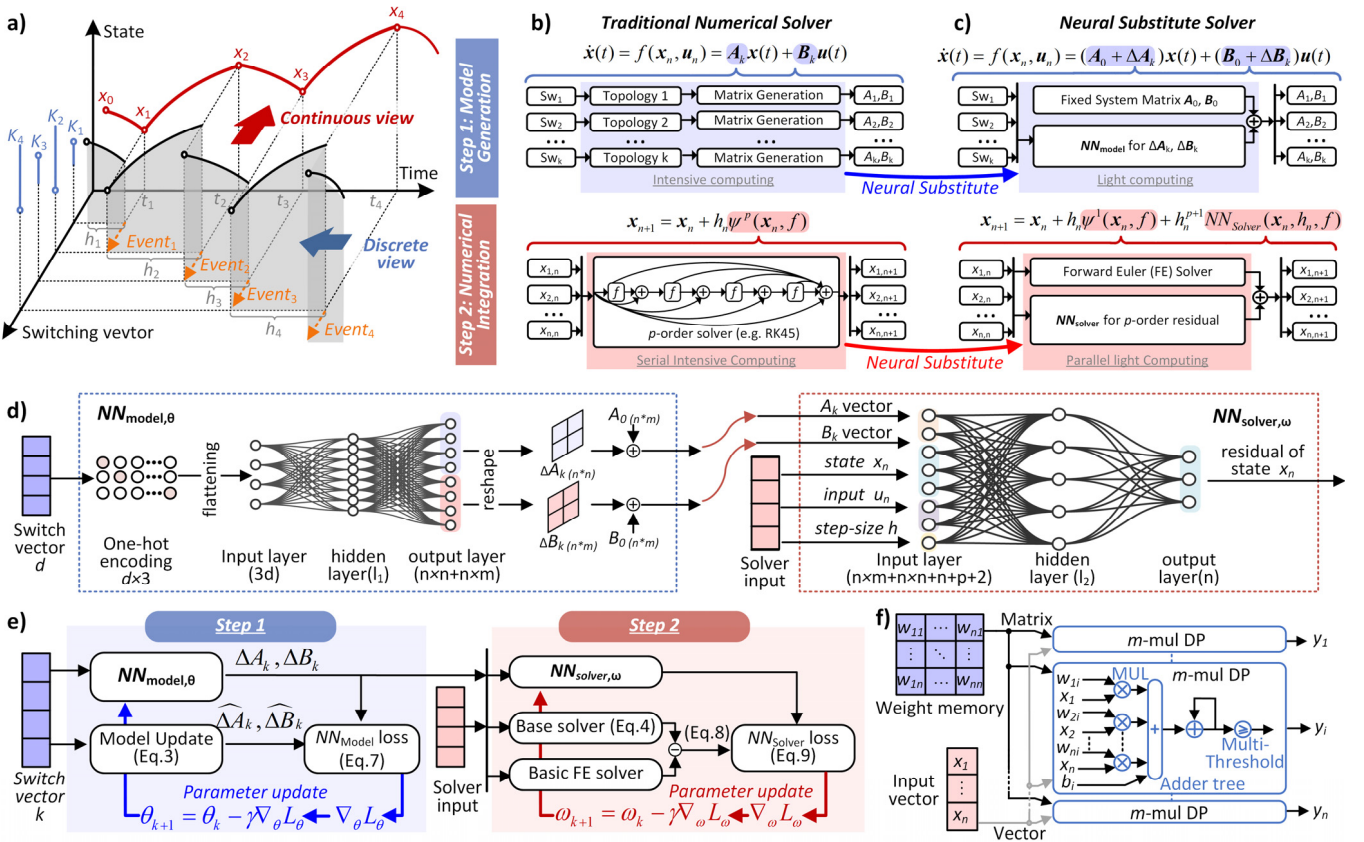

Fig. 1. Schematic diagram illustrating the NSS methodology. (a) Hybrid characteristics of PES. (b) Operation principle of a conventional numerical solver. (c) Operation principle of the proposed NSS. (d) Design details of the NSS. (e) Two-stage training process for the NSS. (f) An FPGA-based neural processing unit tailored for the NSS.

$$\begin{cases} A_k = A_0 + B_1 K_k (I - D_1 K_k)^{-1} C_1 \\ B_k = B_0 + B_1 K_k (I - D_1 K_k)^{-1} C_2 \end{cases} \quad (3)$$

where $A_0$, $B_0$, $B_1$, $C_1$, $C_2$, and $D_1$ are constant matrices derived from the circuit topology, and $K_k \in \{-1, 0, 1\}^d$ is the switching vector determined by the switch-pair modeling method in [4]. Notably, calculating the matrix multiplication and inversion presents a computational bottleneck in online inference

### B. Event-Driven Dynamics Inference

Without analytical solutions, numerical integration is essential for inferring the dynamics of $x(t)$. By employing the event-driven solver (EDS) detailed in [5] to locate the event time and triggering conditions, the integration process first identifies the time interval $[t_k, t_{k+1}]$ corresponding to relevant events $k$ and $k+1$. The system model is then updated using $K_k$ associated with the event $k$. In this time interval, the numerical integrator operates under discrete time step from $t_n$ to $t_{n+1}$:

$$x_{n+1} \approx x_n + h_n \psi(t_n, x_n, u_n, K_k), \quad t_n \in [t_k, t_{k+1}] \quad (4)$$

where $x_{n+1}$ approximates the state at $t_{n+1} = t_n + h_n$, $h_n$ is the step size, and $\psi$ represents the update function of the chosen integration method (e.g., Euler or Runge-Kutta).

Considering the irregular event time and fast-slow dynamics, variable step-size integration methods are preferred over fixed-step ones for efficiency and robustness, which automatically adjust $h_n$ based on local error estimates (e.g., larger steps during smooth dynamics and smaller steps during rapid changes).

Nonetheless, the computational costs for traditional numerical solver remain substantial, as shown in Fig. 1(b). Firstly, it is time-consuming to evaluate the system dynamics within each step, especially generating $A_k$ and $B_k$ with matrix inversions. Secondly, it requires a lot of integration steps to achieve desired accuracy, which is time-uncertain and executed serially.

### III. METHODOLOGIES: NEURAL SUBSTITUTE SOLVER

To address these challenges, this section introduces a neural substitute solver methodology, as shown in Fig. 1(c). NSS aims to substitute computational bottlenecks with neural networks based on the universal approximation theorem, thereby enhancing simulation efficiency to meet the performance requirements and resource constraints of edge computing.

### A. General Formulation

NSS addresses the interplay between the hybrid dynamics and numerical integration by introducing two specialized neural networks that respectively substitute the computational bottlenecks associated with discrete events and continuous states.

Firstly, the model NN ($NN_{Model}$) is employed to substitute the online generation process of the system matrix components ($\Delta A_k$ and $\Delta B_k$) that depend on $K_k$. It learns the mapping from $K_k$ to these matrix variations. Consequently, the system dynamics model actually used in the simulation is approximated by

$$\dot{x} \approx (A_0 + \Delta A_k)x + (B_0 + \Delta B_k)u \quad (5)$$

where $(\Delta A_k, \Delta B_k) = NN_{Model}(K_k)$. It converts time-consuming matrix operation into relatively fast forward propagations.

Secondly, the solver NN ($NN_{Solver}$) enhances the performance of the base variable-step numerical integrator by approximating higher-order error terms. This network learns and substitutes the principal local truncation error term of the base solver (assumed to be $p$-th order), leading to an enhanced single-step integration formula.

$$x_{n+1} \approx x_n + h_n \psi(\cdots; \Delta A_k, \Delta B_k) + h_n^{p+1} NN_{Solver}(\cdots) \quad (6)$$

where the higher-order correction term provided by $NN_{Solver}$ aims to improve the accuracy of the base solver for a given step size $h_n$ (potentially large). By integrating a base solver with an $NN_{Solver}$, correction, the NSS achieves in a single calculation step an accuracy comparable to that of traditional higher-order solvers, which would typically require multiple internal stages or significantly smaller step sizes. This single-step calculation capability between two events makes the NSS particularly effective for hybrid dynamics of PESs. By combining $NN_{Model}$ and $NN_{Solver}$, the NSS framework can significantly reduce both time and resource consumption, while maintaining accuracy.

### B. Details of NSS

**1) Design of $NN_{Model}$:** The objective of $NN_{Model}$ is to accurately learn the nonlinear mapping from discrete $K_k$ to the matrix variations ($\Delta A_k, \Delta B_k$). As shown in Fig. 1(d), the neural network takes the switching vector $K_k$ as input. To prevent the ordinal nature of inputs from influencing the learning process, each element of $K_k$ undergoes one-hot encoding. The resulting set of one-hot vectors is then concatenated and flattened to form a single input vector for the NN. A Multi-Layer Perceptron (MLP) architecture is typically employed, with its structure tailored to the input/output dimensions and the complexity of the mapping task. The MLP outputs a flattened vector containing the elements of the approximated matrices $\Delta A_k \in \mathbb{R}^{n \times n}$ and $\Delta B_k \in \mathbb{R}^{n \times m}$. This output vector is subsequently reshaped to reconstruct these individual matrix forms. To train $NN_{Model}$, one needs to sample various switching vectors $K^{(i)}$ within the operating range and accurately compute the corresponding true matrix variations ($\Delta A_k, \Delta B_k$) using (3) to form a training dataset. The network parameters are then optimized by minimizing a Mean Squared Error (MSE) loss function based on a matrix norm, such as the Frobenius norm:

$$L_{NN1} = \frac{1}{N} \sum_{k=1}^{N} (\|\Delta A_k - \widehat{\Delta A_k}\|_F^2 + \|\Delta B_k - \widehat{\Delta B_k}\|_F^2) \quad (7)$$

**2) Design of $NN_{solver}$:** The objective of $NN_{solver}$ is to learn the $O(h^{p+1})$ local truncation error term of the base variable-step solver to improve accuracy and allow for larger step sizes. First, a simple explicit variable-step method is chosen as the base solver, e.g., variable-step Euler ($p$=1) or second-order Heun's method ($p$=2) [8], with its single-step update function denoted as $\psi$. The typical input of $NN_{solver}$ includes $t_n$, $x_n$, $u_n$, current $h_n$, and the output of $NN_{Model}$. The MLP is also used for the NN.

Residual fitting method is used in training $NN_{solver}$, which requires well-trained $NN_{Model}$. Then, a high-accuracy adaptive solver (e.g., Dopri [8]) is used to solve the system defined by (6), generating a series of reference state points $x_{ref}(t_n)$ and corresponding actual step sizes $h_n$. Next, for each step ($t_n$, $x_{n,\text{ref}}$, $h_n$) from the reference trajectory, the residual $\mathcal{R}_n$ of the base solver relative to the reference solution is calculated by

$$\mathcal{R}_n = \frac{x_{ref}(t_n) - (x_{n-1,ref} + h_n \psi(t_{n-1}, x_{n-1,ref}, u_{n-1}, K_k))}{h_n^{p+1}} \quad (8)$$

Finally, $NN_{solver}$ is trained by minimizing the MSE between its output and the computed residual $\mathcal{R}_n$:

$$L_{NN2} = \frac{1}{M} \sum_{n=1}^{M} \| \mathcal{R}_n - NN_{solver}(t_{n-1}, x_{n-1,ref}, u_{n-1}, h_n, K_{n-1}) \|_2^2 \quad (9)$$

**3) Training Strategy:** As shown in Fig, 1(e), a sequential training strategy is adopted. Initially, $NN_{Model}$ is trained, and its parameters are subsequently fixed. The output of the trained $NN_{Model}$ is then used to generate the training dataset for $NN_{Solver}$, which simplifies the overall training procedure. Although joint training presents an alternative approach, it may lead to increased training complexity. The Adam optimizer is employed for updating network parameters in both training phases. All training was conducted using the PyTorch framework on an NVIDIA RTX 4070 GPU Server.

### C. Parallel Neural Processing Units for NSS

The NSS replaces variable-intensive operations with fixed NN forward propagation by matrix-vector multiplications. To accelerate NSS inference on FPGA-based edge devices, a parallel Neural Processing Unit (NPU) is designed, as shown in Fig. 1(f). The proposed NPU is applicable to both NNs within the NSS considering their similar structures.

The NN parameter sets ($w$ and $\theta$) obtained by server-side training each comprise a weight matrix $W$ and a bias vector $b$. Exploiting the row-wise independence of $W$, the NPU employs multiple dot-product units to concurrently process multiplications between different rows of $W$ and the input vector. Within each dot-product unit, the distributed parallel processing capabilities of FPGAs are leveraged to parallelize operations on individual vector elements. Finally, a multi-threshold unit implements the activation function. This multi-level parallelism facilitates high-throughput low-latency inference on FPGAs, offers a potential magnitude speedup over sequential execution, and meets real-time edge computing demands. Notably, while this parallel approach trades spatial resources for temporal gains, its overall resource cost remains significantly lower than that of conventional methods like matrix inversion.

## IV. CASE STUDY: NSS IN DC-DC CONVERTER

A two-stage DC-DC converter is investigated to evaluate the efficacy of the NSS. Its first stage is a 25 kHz Dual Active Bridge (DAB), and the second stage comprises four parallel 200 kHz Buck converters. A Xilinx FPGA development board, depicted in Fig. 2(a), serves as the edge computing platform.

The NSS was first trained on a server. The converter has 16 switches (8 switch pairs), resulting in an 8-dimensional switch vector. The training set of $NN_{Model}$ comprises all possible switch combinations ($2^8$=256) and corresponding system matrixes. Hyperparameter optimization yielded a final network architecture of {24, 64, 64, 110} neurons with ReLU activation function. The $NN_{solver}$ was subsequently trained using the Dopri solver as a reference. For each of the 256 unique switch vectors, 2000 initial states and step-size were randomly selected and integrated for 0.2 s. This process generated a training dataset totaling 256×2000=512,000 state trajectories. Hyperparameter optimization resulted in a network architecture of {122, 128, 128, 10} neurons with tanh activation function. For testing purposes, a distinct test set was created using 50 randomly selected switch combinations. For each combination, 100 new initial states and step-size were generated, resulting in 50×100=5,000 test trajectories. Upon completion of testing, the trained NSS was deployed on the edge platform leveraging the Xilinx Vitis and Vivado toolchain. The trained NSS is compared with common fixed-step and variable-step solvers in the following ways.

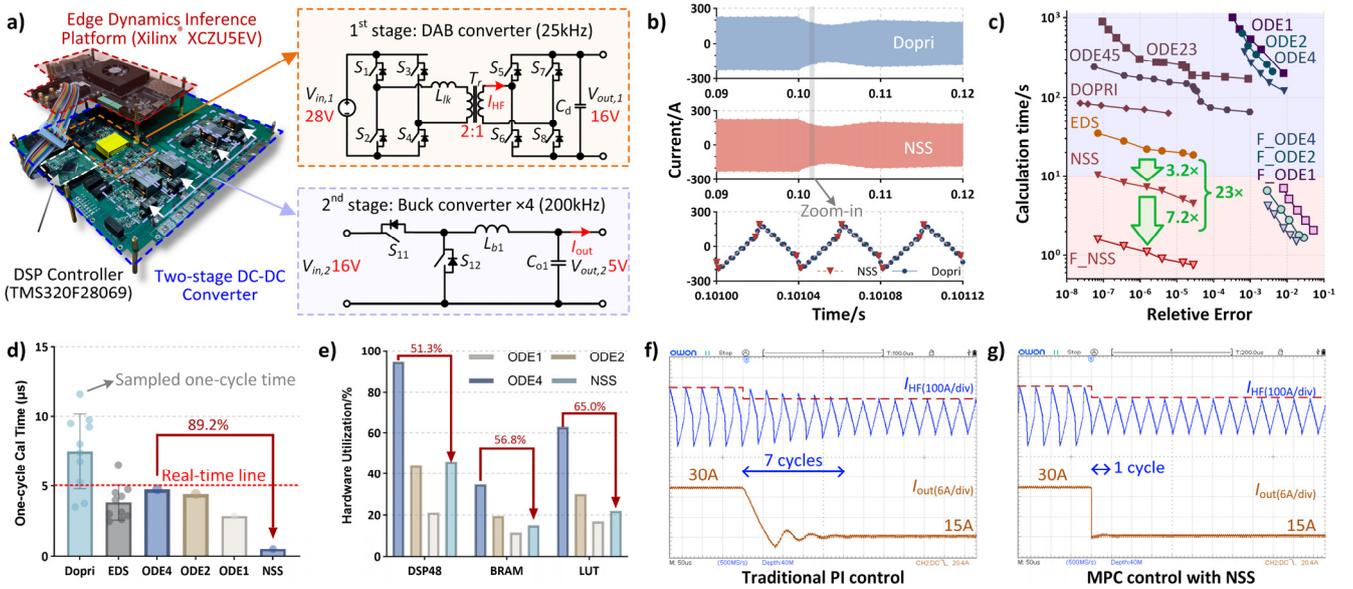

**Fig. 2.** Experimental prototype and comprehensive comparison. (a) Hardware platform and converter topology. (b) Inference accuracy comparison in dynamic scenarios. (c) Solver efficiency comparison for simulating a 10s process. (d) Single-cycle computation time of different solvers on FPGA. (e) Hardware resource utilization of solvers with fixed computation times. (f) Output power regulation using traditional PI control. (g) Output power regulation using MPC with NSS.

**Inference Accuracy:** Unlike steady states used in training, a dynamic process has been employed to evaluate the NSS accuracy in handling switch transitions and numerical integration. As shown in Fig. 2(b), the NSS greatly matches the Dopri solver. Notably, NSS computes only at switching instants, largely reducing computation points.

**Pareto Efficiency:** The efficiency of the NSS and traditional ODE solvers are shown in Fig. 2(c) (the "F" prefix denotes FPGA-accelerated versions). The NSS reduces solution time by 3x when compared to the EDS and by overall 23x with FPGA acceleration, markedly advancing the Pareto efficiency.

**Single-Step Time Stability:** Fig. 2(d) compares the single-step execution time of variable-step solvers capable of real-time operation. Traditional solvers (Dopri, EDS) exhibit fluctuating computation times for timing margins. In contrast, the NSS achieves a constant computation time comparable to fixed-step solvers and is 89.2% lower time than ODE4. This stability is crucial for real-time inference in high-frequency PES.

**Hardware Resource Consumption:** The FPGA resource utilization of three fixed-step solvers and the NSS has been evaluated, as is shown in Figs. 2(e). The NSS shows superior performance in all resource categories and average 60% fewer resources than ODE4.

**Edge Inference Benefits:** For the adjustable power requirements, it is important to improve the dynamic response of its first-stage DAB. Traditional PI control typically exhibits slow transient performance. While an MPC strategy, detailed in [9], can accelerate these dynamics, it necessitates high-bandwidth current sensors and is susceptible to significant communication delays. To show the benefits of edge inference, NSS is integrated with the above MPC strategy. NSS can inference the dynamics of the whole converter, obviating the need for direct high-frequency current sampling by providing high-fidelity current estimates directly to the MPC algorithm. Figs. 2(f) and 2(g) show the experimental dynamic responses of PI control and NSS-MPC, respectively. The NSS-MPC reaches the new steady state within a single switching cycle, validating the accuracy of NSS for edge dynamics inference underscoring its substantial potential for integration within PES.

## V. CONCLUSION

This letter introduces a neural substitute solver for real-time inference of hybrid PES dynamics on resource-constrained edge devices. NSS fundamentally redefines solver application by deeply integrating neural networks to substitute critical computational bottlenecks from traditional methods, strategically shifting this intensive load to offline training for enhanced real-time edge viability. Furthermore, it transforms traditional sequential bottlenecks into highly parallelizable operations, and designs the specific neural processing units based on FPGA hardware for further acceleration. Experimentally, results on a DC-DC converter demonstrate that NSS achieves a 23-fold speedup and a 60% reduction in resource requirements compared to traditional solvers, while also highlighting a significant enhancement to converter control through edge-based dynamics inference. In summary, NSS charts a new path for edge inferring and exploiting PES hybrid dynamics, significantly advancing the paradigm shift of PES development.

## REFERENCES

[1] S. Shao *et al.*, "Modeling and Advanced Control of Dual-Active-Bridge DC–DC Converters: A Review," *IEEE Trans. Power Electron.*, vol. 37, no. 2, pp. 1524–1547, Feb. 2022.

[2] E. Zafra *et al.*, "Computationally Efficient Sphere Decoding Algorithm Based on Artificial Neural Networks for Long-Horizon FCS-MPC," *IEEE Trans. Ind. Electron.*, vol. 71, no. 1, pp. 39–48, Jan. 2024.

[3] D. Gebbran *et al.*, "Cloud and Edge Computing for Smart Management of Power Electronic Converter Fleets" *IEEE Ind. Electron. Mag.*, vol. 17, no. 2, pp. 6–19, Jun. 2023.

[4] J. Zheng *et al.*, "MPSoC-Based Dynamic Adjustable Time-Stepping Scheme for Real-Time HIL Simulation of Power Converters," *IEEE Trans. Transp. Electrification*, vol. 10, no. 2, pp. 3560–3575, Jun. 2024.

[5] J. Zheng, Z. Zhao, Y. Zeng, B. Shi, and Z. Yu, "An Event-Driven Real-Time Simulation for Power Electronics Systems Based on Discrete Hybrid Time-Step Algorithm," *IEEE Trans. Ind. Electron.*, vol. 70, no. 5, pp. 4809–4819, May 2023.

[6] C. Liu, H. Bai, S. Zhuo, X. Zhang, R. Ma, and F. Gao, "Real-Time Simulation of Power Electronic Systems Based on Predictive Behavior," *IEEE Trans. Ind. Electron.*, vol. 67, no. 9, pp. 8044–8053, Sep. 2020.

[7] Z. Li, J. Xu, K. Wang, G. Li, P. Wu, and L. Zhang, "An FPGA-Based Hierarchical Parallel Real-Time Simulation Method for Cascaded Solid-State Transformer," *IEEE Trans. Ind. Electron.*, pp. 1–10, 2022.

[8] Y. L. Kuo and M. L. Liou, "Computer-aided analysis of electronic circuits: Algorithms and computational techniques," *Proc. IEEE*, vol. 65, no. 6, pp. 991–992, Jun. 1977.

[9] S. Wei, Z. Zhao, K. Li, L. Yuan, and W. Wen, "Deadbeat Current Controller for Bidirectional Dual-Active-Bridge Converter Using an Enhanced SPS Modulation Method," *IEEE Trans. Power Electron.*, vol. 36, no. 2, pp. 1274–1279, Feb. 2021.